\documentstyle[11pt,psfig]{article}
\include{psfig}
\def\baselinestretch {1.5}
\evensidemargin 10mm\oddsidemargin 10mm\def\cite#1{#1}
\newcommand{\ct}[1]{[\cite{#1}]}
\def\thebibliography#1{\section*{References}\list
 {[\arabic{enumi}]}{\settowidth\labelwidth{[#1]}\leftmargin\labelwidth
 \advance\leftmargin\labelsep
 \usecounter{enumi}}
 \def\newblock{\hskip .11em plus .33em minus -.07em}
 \sloppy
 \sfcode`\.=1000\relax}

\begin{document}
\begin{center}
{\Large\bf ORIGIN OF THE NUCLEON ELECTROMAGNETIC FORM FACTORS DIPOLE FORMULA.}
\vspace{1mm}
\end{center}
\vspace{0.1cm}

\def\baselinestretch {1}
\begin{center}

{\bf P. Weisenpacher}
\footnote{E-mail address: weisenpacher@center.fmph.uniba.sk}
\\
{\em Department of Theoretical Physics, Comenius University, Mlynsk\'a dolina, 842 48 Bratislava, Slovak
Republic.}\\

\vspace{0.1cm}
\end{center}

\def\baselinestretch {1.5}
\def\sa{\omega}
\def\sbb{{\omega^,}}
\def\sc{{\omega^{,,}}}
\def\sd{{\phi}}
\def\se{{\phi^,}}
\def\va{{\varrho}}
\def\vb{{\varrho^,}}
\def\vc{{\varrho^{,,}}}
\def\vd{{\varrho^{,,,}}}
\def\ve{{\varrho^{,,,,}}}

\def\sam{\omega_0}
\def\sbbm{{\omega^,_0}}
\def\scm{{\omega^{,,}_0}}
\def\sdm{{\phi_0}}
\def\sem{{\phi^,_0}}
\def\vam{{\varrho_0}}
\def\vbm{{\varrho^,_0}}
\def\vcm{{\varrho^{,,}_0}}
\def\vdm{{\varrho^{,,,}_0}}
\def\vem{{\varrho^{,,,,}_0}}

\def\mmm#1{\frac{m_{#1}^2}{m_{#1}^2-t}}
\def\mmt#1#2{\frac{m_{#1}^2m_{#2}^2}{(m_{#1}^2-t)(m_{#2}^2-t)}}
\def\mtt#1#2#3{\frac{m_{#1}^2m_{#2}^2m_{#3}^2}{(m_{#1}^2-t)(m_{#2}^2-t)
     (m_{#3}^2-t)}}
\def\mc#1{{m^2_#1}}
\def\zll#1#2#3#4{{\frac{#1-#2}{#3-#4}}}
\def\zlll#1#2#3#4#5#6#7#8{{\frac{(#1-#2)(#3-#4)}{(#5-#6)(#7-#8)}}}
\def\zzl#1#2#3#4#5#6{{\frac{#1#2}{(#3-#4)(#5-#6)}}}
\def\xxx#1#2{\frac{(#1_N-#1_#2)(#1_N-#1_#2^*)(#1_N-1/#1_#2)(#1_N-1/#1_#2^*)}
               {(#1-#1_#2)(#1-#1_#2^*)(#1-1/#1_#2)(#1-1/#1_#2^*)}}
\def\eee#1#2{\frac{(#1_N-#1_#2)(#1_N-#1_#2^*)(#1_N+#1_#2)(#1_N+#1_#2^*)}
               {(#1-#1_#2)(#1-#1_#2^*)(#1+#1_#2)(#1+#1_#2^*)}}
\def\cc#1{{C_#1}}
\def\zl#1#2#3#4#5{{\frac{#2^#1-#3^#1}{#4^#1-#5^#1}}}
\def\uvv#1{{(\frac{1-#1^2}{1-#1_N^2})}}
\def\fff#1#2{(f^{(#1)}_{#2{NN}}/f_{#2})}
\def\ccc#1#2#3#4#5{\frac{C^{#1}_#2-C^{#1}_#3}{C^{#1}_#4-C^{#1}_#5}}

\def\ffp#1#2{(f^{(#1)}_{#2{pp}}/f_{#2})}
\def\ffn#1#2{(f^{(#1)}_{#2{nn}}/f_{#2})}
\def\ries#1#2#3{\frac{#1}{#2-#3}}

\def\mcsa{{\mc\sa}}
\def\mcsb{{\mc\sbb}}
\def\mcsd{{\mc\sd}}
\def\mcva{{\mc\va}}
\def\mcvb{{\mc\vb}}
\def\mcvc{{\mc\vc}}
\def\mcsc{{\mc\sc}}
\def\mcse{{\mc\se}}
\def\mcvd{{\mc\vd}}
\def\mcve{{\mc\ve}}

\def\ccsa{{\cc\sa}}
\def\ccsb{{\cc\sbb}}
\def\ccsd{{\cc\sd}}
\def\ccva{{\cc\va}}
\def\ccvb{{\cc\vb}}
\def\ccvc{{\cc\vc}}
\def\ccsc{{\cc\sc}}
\def\ccse{{\cc\se}}
\def\ccvd{{\cc\vd}}
\def\ccve{{\cc\ve}}

\begin{abstract}
Starting with the VMD parametrization of the electric and magnetic
nucleon form factors, which are saturated just by the ground state
vector-mesons $\rho$, $\omega$ and $\phi$, then aplying the strict
OZI rule and the asymptotic behaviour of form factors as predicted
by quark model of hadrons, the famous one parameter dipole formula
is derived. By its comparison with space-like data up to $t=-5\
GeV^2$ the most optimal value of the parameter under consideration
is determined. Finaly, charge and magnetization distributions in
proton and neutron are predicted.

\end{abstract}

The electromagnetic structure of nucleons is completely described by four
scalar functions to be dependent on the square momentum transfer of the
virtual photon $t=-q^2$. They can be chosen in the form of Sachs
electric and magnetic form factors
$G_E^p(t)$, $G_M^p(t)$, $G_E^n(t)$ and $G_M^n(t)$. Their behaviour is relatively
complicated, especially in time-like region. In space-like region the
behaviour
of these form factors is well described by so-called dipole formula,
which was founded in 1965
\ct{1} before a discovery of the quark structure of hadrons determining
asymptotic behaviours of the corresponding form factors.
Generally, it has been found
\begin{eqnarray}
G_E^p(t)\approx\frac{G_M^p(t)}{1+\mu_p}\approx\frac{G_M^n(t)}{\mu_n}\approx
\frac{4m_n^2}{t}\frac{G_E^n(t)}{\mu_n}\approx\frac{1}{(1-\frac{t}{0.71})^2},
\label{i1}
\end{eqnarray}
at that time without any theoretical justification.

Since that time theoretical knowledges of form factor behaviours have been
improved. Vector-meson-dominance (VMD) model, based on a experimental fact
of a creation of vector-meson resonances in $e^+e^-$ annihilation into hadrons,
has been created. In the framework of the perturbative QCD quark number
dependent
asymptotic behaviour of form factors has been revealed to be of the form
\begin{eqnarray}
F\sim t^{1-n_q}.\label{i2}
\end{eqnarray}
In this paper using the abovementioned theoretical knowledges the nucleon form
factors dipole formula is derived.

With this aim we start with the canonical VMD parametrization of the electric
and magnetic nucleon form factors
\begin{eqnarray}
\nonumber& &G_E^p(t)=\mmm\va \ffp{E}{\va}+\mmm\sa \ffp{E}{\sa}+\mmm\sd \ffp{E}{\sd}\\
\nonumber& &G_M^p(t)=\mmm\va \ffp{M}{\va}+\mmm\sa \ffp{M}{\sa}+\mmm\sd \ffp{M}{\sd}\\
\nonumber& &G_E^n(t)=\mmm\va \ffn{E}{\va}+\mmm\sa \ffn{E}{\sa}+\mmm\sd \ffn{E}{\sd}\\
& &G_M^n(t)=\mmm\va \ffn{M}{\va}+\mmm\sa \ffn{M}{\sa}+\mmm\sd \ffn{M}{\sd},\label{d1}
\end{eqnarray}
where $f^{(E,M)}_{vNN}$ are vector meson to nucleon coupling constants,
$f_v$ is universal vector-meson coupling constant and
$t$ is photon four-momentum tranfer squared. Taking into account
OZI rule \ct{2,3,4} strictly we require coupling constant of $\phi$
-meson to be zero.

Form factors $G_E^p$, $G_M^p$, $G_E^n$, $G_M^n$ are normalized for the value
$t=0$ as follows
\begin{eqnarray}
G_E^p(0)=1;\;\;G_M^p(0)=1+\mu_p;\;\;G_E^n(0)=0;\;\;G_M^n(0)=\mu_n,\label{d2}
\end{eqnarray}
where $\mu_p$ and $\mu_n$ are anomalous magnetic moments of proton and neutron,
respectively.

For very large space-like $t$ the asymptotic behaviour
\begin{eqnarray}
G_{E,M}^{p,n}\sim t^{-2},\label{d3}
\end{eqnarray}
is applied as predicted by the quark counting rules \ct{5,6}.

Requirements (\ref{d2}) and (\ref{d3}) lead to four systems of
algebraic equations for coupling ratios
\begin{eqnarray}
\nonumber {\rm I}. & &\ffp{E}{\sa}+\ffp{E}{\va}=1\\
\nonumber\  & &\mc\sa \ffp{E}{\sa}+\mc\va \ffp{E}{\va}=0\\
\nonumber\ & &\\
\nonumber {\rm II}. & &\ffp{M}{\sa}+\ffp{M}{\va}=1+\mu_p\\
\nonumber\  & &\mc\sa \ffp{M}{\sa}+\mc\va \ffp{M}{\va}=0\\
\nonumber\ & &\\
\nonumber {\rm III}. & &\ffn{E}{\sa}+\ffn{E}{\va}=0\\
\nonumber\  & &\mc\sa \ffn{E}{\sa}+\mc\va \ffn{E}{\va}=0\\
\nonumber\ & &\\
\nonumber {\rm IV}. & &\ffn{M}{\sa}+\ffn{M}{\va}=\mu_n\\
\  & &\mc\sa \ffn{M}{\sa}+\mc\va \ffn{M}{\va}=0.\label{d4}
\end{eqnarray}

Their solutions take the form
\begin{eqnarray}
\nonumber {\rm I}. & &\ffp{E}{\sa}=-\ries\mcva\mcsa\mcva\\
\nonumber\  & &\ffp{E}{\va}=\ries\mcsa\mcsa\mcva\\
\nonumber\ & &\\
\nonumber {\rm II}. & &\ffp{M}{\sa}=-(1+\mu_p)\ries\mcva\mcsa\mcva\\
\nonumber\  & &\ffp{M}{\va}=(1+\mu_p)\ries\mcsa\mcsa\mcva\\
\nonumber\ & &\\
\nonumber {\rm III}. & &\ffn{E}{\sa}=0\\
\nonumber\  & &\ffn{E}{\va}=0\\
\nonumber\ & &\\
\nonumber {\rm IV}. & &\ffn{M}{\sa}=-\mu_n\ries\mcva\mcsa\mcva\\
\  & &\ffn{M}{\va}=\mu_n\ries\mcsa\mcsa\mcva,
\end{eqnarray}

which tranform form factors defined by (\ref{d1}) into relations
\begin{eqnarray}
\nonumber& &G_E^p(t)=\frac{1}{(1-\frac{t}{\mcsa})(1-\frac{t}{\mcva})}\\
\nonumber& &G_M^p(t)=(1+\mu_p)\frac{1}{(1-\frac{t}{\mcsa})(1-\frac{t}{\mcva})}\\
\nonumber& &G_E^n(t)=0\\
& &G_M^n(t)=\mu_n\frac{1}{(1-\frac{t}{\mcsa})(1-\frac{t}{\mcva})}.\label{d5}
\end{eqnarray}

Masses of $\omega$ a $\rho$ mesons are almost the same, therefore
one can substitute these values in relations (\ref{d5}) by
averaged mass $m$ and finally for electric a magnetic nucleon form
factors one gets expressions
\begin{eqnarray}
\nonumber& &G_E^p(t)=\frac{1}{(1-\frac{t}{m^2})^2}\\
\nonumber& &G_M^p(t)=(1+\mu_p)\frac{1}{(1-\frac{t}{m^2})^2}\\
\nonumber& &G_E^n(t)=0\\
& &G_M^n(t)=\mu_n\frac{1}{(1-\frac{t}{m^2})^2}\label{d6}
\end{eqnarray}
consistent with the standard form of dipole formula besides the
form factor $G^E_n(t)$, which is equal zero. This fact correspond
with experimentally detected negligible values of $G^E_n(t)$.
Dipole formula constant value is approximately $m^2=0.60\
GeV^{2}$.

Behaviour of form factor $G_E^n(t)$ can be determined including their
additional properties. One can start from the identity
\begin{eqnarray}
<r_n^2>=6\frac{dG^n_E(t)}{dt}\mid_{t=0}=6\frac{dF_1^n(t)}{dt}\mid_{t=0}+
\frac{3\mu_n}{2m_n^2},\label{d7}
\end{eqnarray}
following from a decomposition of $G_E^n$ into Dirac a Pauli form factors
\begin{eqnarray}
\nonumber G^n_E(t)=F_1^n(t)+\frac{t}{4m_n^2}F_2^n(t)
\end{eqnarray}
and normalization condition $F_2^n(0)=\mu_n$. According to the
fact, that the last term value in (\ref{d7}) $-0.126\ fm^2$ is
comparable with experimentally determined value of the neutron
charge radius ($<r_n^2>_{exp}=-0.119\ fm^2$ \ct{7}), the term
$\frac{dF_1^n(t)}{dt}\mid_{t=0}$ is almost zero and it can be
neglected. So, taking into account the later and the zero value of
$G_E^n(t)$ at $t=0$ one obtains a dipole formula for electric
neutron form factor as follows
\begin{eqnarray}
G_E^n(t)=\frac{\mu_n}{4m_n^2}t\frac{1}{(1-\frac{t}{m^2})^2}\label{d8}
\end{eqnarray}
which describes data quite well.

If we define
\begin{eqnarray}
G_D(t)=\frac{1}{(1-\frac{t}{\lambda})^2},\label{d9}
\end{eqnarray}
then the electric and magnetic nucleon form factors take the form
\begin{eqnarray}
\nonumber& &G_E^p(t)=G_D(t)\\
\nonumber& &G_M^p(t)=(1+\mu_p)G_D(t)\\
\nonumber& &G_E^n(t)=\frac{\mu_n}{4m_n^2}tG_D(t)\\
& &G_M^n(t)=\mu_nG_D(t),\label{d10}
\end{eqnarray}
identical with (\ref{i1}).

There are 433 experimental values of form factors in space-like region
obtained from the elastic electron
scattering on hydrogen and deuteron target, which have been
analyzed by means of the
relations (\ref{d10}) using program MINUIT.

One could expect, that two free parameter dependent dipole formula in
(\ref{d5}) gives a better
description of experimental data than canonical one free parameter
dipole formula (\ref{d10}). However, the values of both free parameters
in (\ref{d5}), determined in
fitting procedure take almost the same values and
therefore (\ref{d5}) is practically reduced to
(\ref{d6}). In this case the dipole formula parameter takes the value
$\lambda=0.7134\ GeV^{2}$ ($\chi^2/ndf=6.32$). Introducing non-zero value of
$G_E^n(t)$ by (\ref{d8})
we obtain better description $\lambda=0.7132\ GeV^{2}$
($\chi^2/ndf=5.92$).

In order to achieve acceptable value of $\chi^2/ndf$ it is
necessary to reduce our fitting procedure to points up to $-5\
GeV^2$. $\chi^2/ndf$ takes a value $2.34$ and $\lambda=0.7263$.
This results are graphically presented in Fig. 1. Strong
enhancement of $\chi^2/ndf$ value suggests important deviation
from dipole behaviour for low values of four-momentum transfer.
Approximately 20\% of total $\chi^2$ is generated by $G_E^n(t)$.
Relation (\ref{d8}) gives a worse description of experimental
data, therefore it is impossible to decrease $\chi^2$ value per
point below limit $\chi^2/ndf=2$.

Taking into account (\ref{d10}) one can predict charge and
magnetization distribution of nucleons (Fig. 2). These results are
compared with distributions given by unitary and analytic
ten-resonance model presented by \ct{8}.

Although dipole formula gives a good description of all
experimental values of electric and magnetic nucleon form factors,
since this time it has not been known any theoretical motivation
of these relations. Our paper present derivation of dipole formula
including VMD model saturated by ground state of vector-meson
$\rho$, $\omega$ and $\phi$, OZI rule, normalization conditions
and asymptotic behaviour of form factors given by quark model of
hadrons.

I would like to thank Professor Stanislav Dubni\v cka for
suggesting this problem and for his giving me new ideas that led
to successful solution, and Professor Anna Zuzana Dubni\v ckov\'a
for a great support and useful advice.

\begin{figure}[t]
\vspace{-1.8cm} \centerline{\psfig{figure=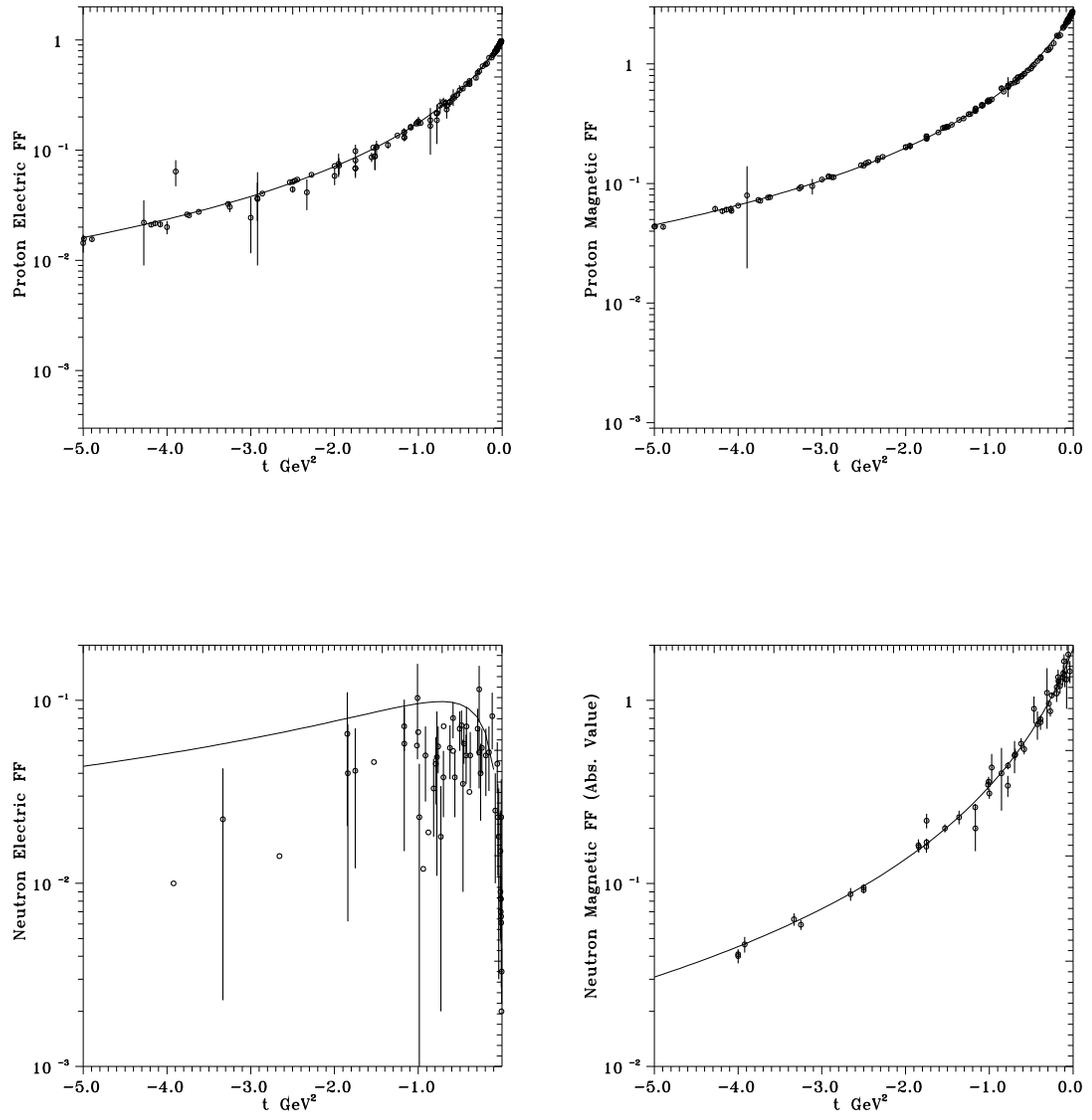,height=15.cm}}
\vspace{1.0cm} \caption{Behaviour of nucleon electric and magnetic
form factors in space-like region given by dipole formula.}
\label{Fig.1}
\end{figure}

\begin{figure}[t]
\vspace{-1.8cm} \centerline{\psfig{figure=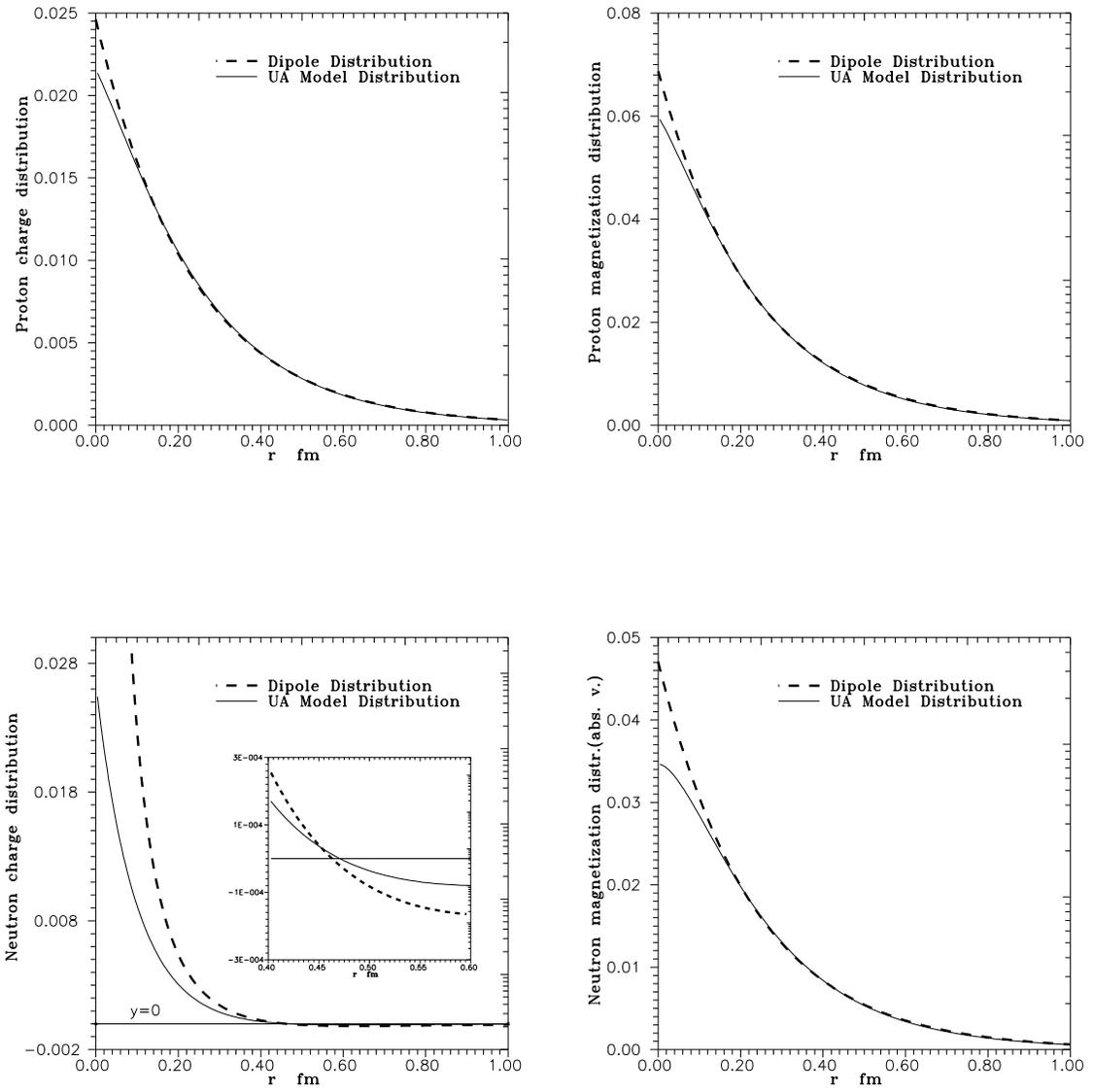,height=15.cm}}
\vspace{1.0cm} \caption{Predicted charge and magnetization
distribution in proton and neutron.} \label{Fig.2}
\end{figure}

\end{document}